# An Optimization-Based Model for Full-body Reaching Movements

Daohang Sha[1]*, James S Thomas[2]*[§]

[1]Department of Orthopaedic Surgery, Yale Medical School, New Haven, CT, USA

[2]School of Physical Therapy, Ohio University, Athens, OH, USA

*These authors contributed equally to this work

[§]Corresponding author

Email addresses:

    DS: sha@ohiou.edu

    JST: thomasj5@ohiou.edu

# Abstract

**Background**

The development of a simulation model of full body reaching tasks that can predict end-effector trajectories and joint excursions consistent with experimental data is a non-trivial task. Because of the kinematic redundancy inherent in these multi-joint tasks there are an infinite number of postures that could be adopted to complete them. By developing models to simulate full-body reaching movements in 3D space we can begin to explore cost functions that may be used by the central nervous system to plan and execute these movements.

**Methods**

A robust simulation model was developed using 1) graphic-based modeling tools to generate an inverse dynamics controller (SimMechanics), 2) controller parameterization methods, and 3) cost function criteria. An adaptive weight coefficient based on the final motor task error (i.e. distance between end-effector and target at the end of movement) was proposed to balance motor task error and physiological cost terms (e.g. joint power). The output of the simulation models using different cost controller functions based on motor task error or motor task error and various physiological cost terms (e.g. joint power, center of mass displacement) were compared to experimental data from 15 healthy participants performing full body reaching movements.

**Results**

In sum, the best fit to the experimental data was obtained by minimizing motor task error, joint power, and center of mass displacement. Simulation and experimental results demonstrated that the proposed method is effective for the simulation of large-scale human skeletal systems.

**Conclusions**

This method can reasonably predict the whole body reaching movements including final postures, joint power and movement of COM using simple algebraic calculations of inverse dynamics and forward kinematics.

# Background

Full body reaching tasks require the central nervous system (CNS) to apportion motion to the legs, trunk and arms in order to successfully complete the task. The two primary constraints are that the end-effector makes contact with the target, and that the body's center of mass remains within the base of support (i.e. at target contact). However, even with these constraints, there are an infinite number of joint configurations that can be used to complete the task due to the inherent kinematic redundancy in the human body. Given an infinite solution set, how does the CNS plan and execute coordinated movements in a kinematics redundant system. From a mathematical perspective, optimal control theory is an effective method used to solve redundant systems. In fact some have suggested that the CNS uses a method similar to optimal control to coordinate multi-joint movements [1, 2].

In optimal control, the apportionment of motion to the various joints is determined by an iterative process that attempts to minimize certain input criteria. Various criteria for the optimal control of human motor coordination have been proposed based on empirical findings, physiological phenomena or both. Flash and Hogan (1985) reported that smoothness of the end-effector trajectory in Cartesian space is optimized (i.e minimum end-effector jerk) [3], while Rosenbaum *et al*. (1985) found that smoothness of joint space is optimized (i.e. minimum joint angular jerk) [4]. Still others have proposed that joint torque [5], joint power [6]; or displacement of whole body center of mass (COM) [7] is minimized. Finally, Nakano *et al*, 1999 proposed that final end-effector error is minimized as well as norm of torque change (i.e. the norm of $1^{st}$ order derivative of joint torque) [8].

In motor control simulations, the goal is to determine the apportionment of joint motions based on some optimized criteria (e.g. minimal joint torque) and this requires that the optimal controller input joint torques or muscle forces to produce joint motions [5, 9, 10]. This process is called forward or direct dynamics. This method requires an integration of the differential equations of the multi-body systems, which can result in the entire algorithm being unstable, particularly when the gravitational and elastic forces are included in the dynamics model. While several methods have been proposed to improve the stability of these forward models [2, 9], they are still problematic in terms of the computation time required to perform these calculations, i.e. up to 10,000 hours of CPU time in desktop computers [5, 11]. This is because initial conditions must be checked for consistency with constraints after each of iteration of integration.

An alternative approach uses an inverse dynamics model in which the optimal controller computes joint angular positions derived from a certain criterion. These joint angles are then used as input in an inverse dynamics model that computes the joint torques or internal forces [6, 12]. The inverse dynamics based method is much faster than the forward dynamics based method and eliminates the stability problem for the system dynamics due to its intrinsic algebraic structure [6]. However, a potential drawback of this method is that the inverse dynamics model only works on open topologies, i.e. open link chains[13].

While there are several ways to implement a dynamics system model, once the system's degrees-of-freedom (DoF) increases, the required analytical expressions become unwieldy [2]. While these expressions could be determined using symbolic tools, a more efficient method is to use graphic-based tools for simulation of rigid body machines such as SimMechanics. SimMechanics requires the geometry of bodies and mass properties, possible motions, kinematics constraints, and the coordinate systems to initiate the model. It doesn't require the user to develop the equations of motion independently [14]. Additionally, graphic tools allow real-time visualization during simulation, which is useful to validate the proposed model because the simulated movement should at least be qualitatively similar to what is observed experimentally [11].

The purpose of this study is to develop an effective method/model to address how motor system performs full body reaching movements. In this paper, we present an optimal controller based on the parameterization optimization method using several different input criteria that satisfy the constraints of a full body reaching task. A method for using graphic based tools based on inverse dynamics to model full body reaching tasks is presented. Finally, the models are validated qualitatively and quantitatively to experimental data collected from healthy subjects performing full body reaching tasks.

## Methods
**Simulation Model**

In voluntary target reaching activities, the primary goal of the motor task is to ensure that the end-effector makes contact with the target. As illustrated in Figure 1, the first problem for the control system is to complete the task with minimal end-effector error (task error). Cost functions generate parameters that are input to the polynomial controller, which outputs a set of joint angle trajectories. These joint angles are input to the inverse dynamics and forward kinematics models, which in turn provide output of joint torques, power, COM location and end-effector error. These are then used iteratively to compare with task constraint (i.e. end-effector reaching the target location) to refine the cost function. The cost functions are then minimized iteratively using an optimization algorithm (Figure 1). Each of the blocks of our model, as illustrated in Figure 1, is described in greater detail below.

**Cost function**

In general, the criteria or cost functions for voluntary target reaching movement can be written as 
$$C = e_f^T e_f + \lambda \int_0^{t_f} u^T R u \, dt \quad (1)$$

where $e_f$ is the final motor task error / end-effector error vector with three elements representing errors in *x*, *y*, and *z* direction respectively (i.e. distance between end-effector and target at the end of movement); $\lambda$ is a weighting coefficient that expresses the relative importance between the motor task error (first term) and the physiological cost (second term); *R* is a positive-definite matrix with proper dimensions, which indicates the importance of each joint involved in the full body reaching; *u* could be one of the following quantities or their combinations such as end-effector jerk $\frac{d^3x}{dt^3}$, joint torque $\tau$, joint torque change $\frac{d\tau}{dt}$, joint power $P = \tau \times \dot{\theta}$, or body center of mass $x_C = \frac{\sum m_i x_i}{\sum m_i}$. All of these quantities are the function of parameters of the controller (see next section). It must be emphasized that all calculations of physiological cost functions include the effect of accelerations due to gravity.

As presented above, the cost function has an end-effector error term and a physiological cost term. In this paper physiological cost refers to movement performance measures such as total joint power, displacement of center of mass, or a combination of these two variables. The first term and second term are usually in different units (e.g. m$^2$ and (N.m)$^2$ ) and depending on the physiological cost term, their values can be vastly different. Because this is a multiple objective optimization which requires the simultaneous optimization of more than one cost function, some trade-off between the criteria is needed to ensure a satisfactory movement prediction. Here we propose an adaptive weight coefficient that adjusts its value based on the final motor task error, i.e. $\lambda = \lambda_0 e_f^T e_f$ with constant $\lambda_0$. So the criterion becomes

$$C = e_f^T e_f \left[ 1 + \lambda_0 \int_0^{t_f} u^T R u \, dt \right] \tag{2}$$

Now the constant $\lambda_0$ can be chosen so the value of the physiological cost term is equal to one when the motor task error satisfies the preset tolerance. When the two terms in brackets are approximately equal then optimization convergence is assured. The constant can be calculated by $\lambda_0 = \left( \max \int_0^{t_f} u^T R u \, dt \right)^{-1}$. The maximum value of integration in the equation can be obtained by the primary simulation with end-effector error cost as the only criterion. Once the primary simulation has been run, and then $\lambda_0$ can be determined for each physiological cost term (e.g. joint power, COM). This ensures the end-effector reaches the target location at the end of movement if the optimization converges. This is a necessary though not sufficient condition for convergence. Actually, once the cost function is determined, the convergence depends mainly on the behavior of the optimization method.

**Polynomial Type Controller**

The time history of the optimal trajectory of each joint is a function of time which can be approximated by an $n^{th}$-order polynomial

$$\theta(t) = p_0 + p_1 t + p_2 t^2 + p_3 t^3 + \ldots + p_n t^n. \tag{3}$$

The angular velocity and acceleration can be derived analytically from equation (3). These joint trajectories are then used as input for the inverse dynamics and forward kinematics calculations (Fig.1). The advantage of using an inverse dynamics method is that these calculations do not require integration, and since there is a one-to-one mapping from joint space to Cartesian space in forward kinematics, the problem of kinematic redundancy is eliminated. Thus, by using a forward kinematics calculation, the problem is reduced to determining the coefficients of the polynomial. Furthermore, the whole body voluntary reaching movement can be partially described as the point where the end-effector starts ($t = 0$) moving (i.e. standing neutral posture where $\theta_0$ is known) to the posture adopted at contact target within a certain time ($t_f$). Thus, the location of target in Cartesian space is known but the final posture $\theta_f$ is not.

Based on equation (2), with the initial conditions $\theta(0) = \theta_0$, $\theta(t_f) = \theta_f$, $\dot{\theta}(0) = 0$, and final conditions $\dot{\theta}(t_f) = 0$, $\ddot{\theta}(0) = 0$, $\ddot{\theta}(t_f) = 0$,

and using a 6$^{th}$ order polynomial as example, one gets the following:

$$p_0 = \theta_0,$$
$$p_1 = 0, \quad \text{and} \quad$$
$$p_2 = 0,$$

$$p_3 = \frac{1}{t_f^3}\left[10(\theta_f - \theta_0) - p_6 t_f^6\right],$$
$$p_4 = \frac{1}{t_f^4}\left[-15(\theta_f - \theta_0) + 3 p_6 t_f^6\right], \quad (4)$$
$$p_5 = \frac{1}{t_f^5}\left[6(\theta_f - \theta_0) - 3 p_6 t_f^6\right]$$

This method requires at least a 6$^{th}$ order polynomial to ensure enough freedom for the approximation of joint movements as is evident from above. If the final posture is unknown, the variable $\theta_f$ can be tuned together with other parameters such as $p_6$. Therefore, there are only two variables (i.e. final posture $\theta_f$ and polynomial parameter $p_6$) that need to be tuned for each movement of each joint. The higher the order of polynomial, the more freedom the polynomial has to approximate the joint motion, however, as the polynomial order increases the number of variables that needs to be tuned also increases.

**Skeleton Dynamics**

The general equations of full body motion can be written as

$$\tau = I(\theta)\ddot{\theta} + V(\theta,\dot{\theta}) + G(\theta)g + T(t) \quad (5)$$

where $\theta, \dot{\theta}, \ddot{\theta}$ are the vectors of joint angle, angular velocity, angular acceleration respectively; $I(\theta)$ is segment mass inertia moment matrix; $\tau$ is vector of net joint moment; $G(\theta)$, $V(\theta,\dot{\theta})$, $T(t)$ are gravity terms, Coriolis-centripetal-viscoelasticity, and external terms such as ground reaction forces. Joint viscoelasticity can be written as $V(\theta,\dot{\theta}) = K\theta + B\dot{\theta}$, where $K$ and $B$ are joint stiffness and viscoelastic coefficient matrices respectively.

One way to implement the dynamics system model is to use analytical methods. For systems with low DoF (i.e. DoF<3), the analytical expressions for the relationships between angular accelerations and joint torques can be easily written. However, a system with as few as two segments with 7 DoF, requires an analytical expression with more than 200,000 elementary operations (e.g. +,-,*, cos, sin) [2]. Although these expressions could be determined using symbolic tools, it is still unimaginable that how many elementary operations would be needed for a full body model with 12 segments and 36 DoFs. Therefore, a more efficient method is to use graphic-based tools for simulation of rigid body machines such as SimMechanics (The MathWorks, Inc.) or SIMM (Musculaographics, Inc.).

In this paper, a linked segment model for the inverse dynamics of whole body motions in 3D space was developed using Matlab/Simulink and SimMechanics Toolbox. This linked segment model includes twelve segments (i.e. head, thorax, abdomen, pelvis, right/left hands, forearms, upper arms, thigh, and shank) and twelve joints (i.e. right/left wrist, elbow, shoulder, cervical, thoracic, lumbar, hip, knee and ankle). While this model was developed using only one leg because of inherent problems with solving inverse dynamics of closed loop systems using SimMechanics, the movements of the lower extremities are small and nearly symmetrical. Thus this single leg model should provide reasonable results. To further simplify the model, each joint has only three rotational degrees of freedom (DoF), (i.e. flexion/extension, internal/external rotation and abduction/adduction) within the sagittal, frontal and transverse planes. The total number of DoF for the model is 36. While others have included translation of the shoulder girdles in their models [15], we chose to use a more simplified model that could be readily compared to our experimental data. The inputs for the model (i.e. the joint angular trajectories and their derivatives) are provided by the polynomial controller. The outputs of the model (computed by SimMechanics), are the net muscle torques and forces for each joint, as well as joint power and motion of COM, end-effector location $x = g(\theta)$ (forward kinematics). These outputs then are used to reduce end-effector error and physiological cost through the use of an optimization algorithm.

**Optimization Algorithm**

The cost function is minimized subject to the equality constraint (nonlinear dynamics equation of motion, Eq.5; initial and final boundary conditions, Eq.4) and the inequality constraints (the limitation of joints) to obtain the optimal parameters of controllers, i.e.

$$p_{opt} = \arg \min_{\|\Delta p\| \leq \varepsilon_1, C \leq \varepsilon_2} C(p) \quad \text{subject to} \quad \theta_i^{low} \leq \theta_i \leq \theta_i^{up}, i = 1,...,n$$

where $p$ is a parameter vector of controllers; therefore the cost function is also a function of the unknown coefficients $p_i, i = 1,...,n$ and the final time $t_f$; $\|\Delta p\| \leq \varepsilon_1, C \leq \varepsilon_2$ are the stop conditions of the optimal algorithm.

Once the initial values for the parameters of controllers have been input, the optimization algorithm will modify the parameters until the preset criteria or minimum parameter changes are satisfied, i.e. $p^{j+1} = p^j + \Delta p^j$, where $j$ is the index of iteration. In principle, any well developed nonlinear optimization algorithm can be used to find the optimal controller parameters. However, different algorithms will produce different results. Here we use the nonlinear Least-squares function `lsqnonlin` in Matlab Optimal ToolBox to perform the optimal parameter search. The Levenberg-Marquardt method with line search was used (More 1977), i.e.

$$\Delta p^j = -\alpha^j \left[ J^T J + \sigma^j I \right]^{-1} \nabla C(p^j), \tag{7}$$

- 11 -

where $J = \left[ \dfrac{\partial \sqrt{C_t}}{\partial p_i} \right]$ with $t = 0, \ldots, t_f$ and $i = 1, \ldots, n$, is Jacobian matrix; $\alpha^j$ is the iteration step which is determined by line search; $\sigma^j > 0$ is a positive constant; $\nabla C(p^j) = \dfrac{\partial C(p^j)}{\partial p^j}$ is the gradient of criterion $C$ respective to controller parameter $p$. Jacobian matrix ($J$) and the gradient of criterion ($\nabla C(p^j)$) are approximately calculated through parameter perturbation.

**Initializing Simulation Model**

As noted earlier the cost function is a composite function with a task error cost term and a physiological cost term. Simulations were conducted using the following criteria cost functions: 1) minimize final end-effector error only without any physiological cost term (i.e. min Error), 2) minimize final end-effect error and total joint power (i.e. min Power), 3) minimize final end-effector error and body COM displacement (i.e. min COM), 4) minimize final end-effector error, total joint power and body COM displacement (i.e. min Power+COM).  We also calculated the hand trajectory using minimal jerk criteria (min Jerk) as described by Flash and Hogan (1985) [3]. The minimum-jerk trajectory of end-effector was calculated by the following equation,

$$x(t) = x_0 + (x_f - x_0)\left[ 10\left(\dfrac{t}{T}\right)^3 - 15\left(\dfrac{t}{T}\right)^4 + 6\left(\dfrac{t}{T}\right)^5 \right], \tag{8}$$

when moving from location $x(0) = x_0$ to $x(T) = x_f$ in $t = T$ seconds [3].

To initialize the optimization, the initial joint angles at ($t=0$) and ($t=t_f$) were set to the mean values from neutral standing posture derived from experimental data, and the controller parameters were set to zero.  In this study, a 6$^{th}$ order polynomial controller with two unknown variables, i.e. final posture ($\theta_f$) and the coefficient of polynomial ($p_6$) is used to determine each joint excursion. Therefore, there are 72 parameters to be tuned for the 12 joints. To ensure physiological fidelity of the simulation, joint range of motion values (Eq.3) were input based on accepted norms of joint range of motion [16], and measures of joint viscoelasticity determined from the extant literature [17-20] (See Table 1).

The average height and weight of 15 healthy subjects were used in the simulations and mass-inertial characteristics of each segment were derived from the regression equations provided in the literature [21, 22].

Final motor task error (i.e. distance between target and end-effector), parameter tolerance ($\varepsilon_1$) and criterion tolerance ($\varepsilon_2$) of termination of optimization algorithm were set as 2mm, $10^{-6}$, and $10^{-6}$ respectively for all simulations. The constant weight $\lambda_0$ was set as $10^{-8}$, $10^{-9}$, $10^{-10}$ (for high, middle and low targets) and $7\times10^3$, $10^3$, $10^2$ (for high, middle and low targets) for power and COM term respectively. For simplicity, joint weight matrix $R$ was set as the unit matrix. The movement duration was set as 0.56s, 0.575s and 0.68s for high, middle and low targets respectively based on the experimental results from 15 healthy subjects. The solver for numerical computation was set with a fixed-step sample interval of 0.001s.

**Experimental Protocols**

Fifteen healthy subjects (7 males and 8 females with age 22.93 ± 1.79 year, weight 68.59 ± 10.69 kg and height 169.12 ± 7.74 cm) performed a series of reaching tasks to three targets located in the mid-sagittal plane at a fast paced speed (i.e. approximately 600 ms from initial posture to target contact). Target locations were standardized to the participant's anthropometrics. The participants could, in theory, reach the targets by flexing their trunk 15, 30, or 60 degrees, with their shoulder flexed to 90-degrees and elbow extended without any motion from the other joints [23]. Before beginning the study, each participant was informed of the experimental protocol and signed the consent form approved by the Ethics Committee of the Ohio University.

Starting from an upright standing posture, with each foot on a force plate, the participant reached with their right hand for the target. Subjects paused at the target for 1 second and then returned to an upright posture. Three trials at each target height were performed and the targets were presented from highest to lowest. Motions of the trunk and limb segments were recorded at 120Hz for 5 seconds using the MotionMonitor System (Innovative Sports Training, Inc. Chicago). This system can track the three-dimensional coordinates of six-degree-of-freedom magnetic sensors with a spatial resolution of 1.8mm in position and 0.5deg in orientation (Ascension$^{TM}$, Flock of Birds®, Ascension Technology Corporation, Burlington, VT, USA). The magnetic sensors were attached by Velcro® straps to the limb segments (at the midpoint between the joints) of the right and left shank, thigh, arm, and forearm, as well as the thoracic vertebra (T1), lumbar vertebra (L1), and the sacrum. An Euler angle sequence was used to derive the three dimensional joint motions from the upper and lower extremities bilaterally, as well as the thoracic and lumbar spine. These data were smoothed with a 61-point fourth order Savitzky-Golay filter [24] and served as input for inverse dynamics calculations. The same inverse dynamics (Simulink model) model used in simulation was used to calculate the motion of COM, joint torques, and joint power for each trial of each subject. The properties of each segment, such as mass-inertial characteristics, size, and location of COM etc., were derived from anthropometric regression equations based on the mass and height of each subject [21, 22]. The experimental trajectory of end-effector including distance from target location, velocity, acceleration and jerk also were calculated from the same Simlink model.

**Data Analysis**

The paths/trajectories of the end-effector determined from the simulation model were compared qualitatively to the means and standard errors of path/trajectories of end-effector from the experimental data. To quantitatively assess the simulation model, for each target height, t-tests were used to compare the predicted COM displacement and predicted total joint power to the experimental data.

# Results

**Optimization**

The optimization for each trial only takes a few hours of CPU time on a personal computer (see Table 2), however, there was no clear effect of cost functions or target location for the CPU time required to perform the simulations.

Examination of total power squared and final COM squared (i.e total cost) in Table 2 shows that the values of these measurements are significantly reduced when adding the physiological cost terms (e.g. min Power, min COM and min Power + COM) to the Error cost term even though we used an adaptive weight coefficient to balance the physiological cost terms. Comparing all four control strategies, final COM squared was smallest when the min COM strategy was used. However, total power squared was smallest for min Power + COM control strategy.

**End-effector Trajectories**

Figure 2 illustrates the trajectories of the end-effector determined by 1) the simulation model using four different cost functions, 2) calculated minimum jerk, and 3) experimental data averaged over 15 subjects. Visual inspection of this figure reveals that for each target height, the trajectories of the end-effector and their derivatives are quite similar to the experimental data regardless of the cost criteria used in the simulation. The bell shaped velocity traces are consistent with previous findings for two and three joint arm movements. While these trajectory plots suggest that each simulation method provides similar results, examination of the path of the end-effector gives greater insight into the differences in these methods. The path of the end-effector determined by the different simulation methods and the experimental data (as described above) is shown for each plane and target height in Figure 3. Examination of the experimental data for the path of the end-effector in the sagittal and frontal planes (Figure 3A & B) reveals a fairly large path curvature, which is consistent with the path predicted by the simulation using each cost function. In contrast, using minimal jerk as a cost function predicts a straight line path of the end-effector in all planes. However, examination of the experimental data for the path of end-effector in the transverse plane indicates a much straighter path. Thus, for the transverse plane, the best fit appears to be provided by either minimal jerk or the minimal task error cost function. In contrast, minimal power and COM cost functions predict a fairly large curvature and do not provide a good fit of the experimental data. Nonetheless, as stated earlier, even if the path of the end-effector can be accurately predicted, the movement strategy or the posture adopted at target contact may not be consistent with experimental data. Thus, we next compare the predicted COM displacements to the experimental data for each target height.

**Displacement of COM**

The trajectories of COM as predicted by the simulation models using the different cost functions are compared to the trajectories of COM from the experimental data (Figure 4). While, the exact path of the COM does not appear to be well fit by any of the cost functions, the change in COM from initial posture to target contact appears to be consistent with the cost function that minimized joint power and COM displacement. In fact, t-tests revealed that using a cost function that minimized COM displacement and joint power was the only cost function that was not significantly different from the experimental data (Table 3). Minimum task error predicted the largest displacement of the COM. Specifically, for the middle and low target reaching tasks, the predicted COM displacement exceeds 20 cm and 30 cm respectively, which would clearly cause the subject to fall forward or require a step to prevent falling. Therefore minimizing end-effector error alone does not provide a reasonable solution for the movement task with respect to COM displacement. Notice that when only task error is minimized, the posture adopted at target contact does not compensate for the forward displacement of the trunk and its effect on whole body COM (Figure 5). Qualitatively, the posture adopted at target contact by participants (i.e. experimental data) is best fit by the cost function that minimizes COM displacement and joint power (Figure 5).

**Total Joint Power**

Figure 6 shows the time series of joint power $\sum_{i=1}^{36}\left|\tau_i(t)\dot{\theta}_i(t)\right|$ derived from experimental data and from model simulations using the various optimal control strategies. From this figure, in general, it appears that the cost functions of minimal (power + COM) and minimal power give the best qualitative fit of the experimental data for the high target only. However, for the middle and low targets, only minimum (power + COM) appear to provide a good fit of the data. To compare the cost functions to the experimental data quantitatively, we took the integral of joint power to get total energy (See Table 4). For the high target and middle target it appears that minimizing COM displacement provides the best fit to the experimental data. However, for the low targets, none of the cost functions provide a good fit to total energy expenditure.

# Discussion
### Skeleton Dynamics

The underlying premise of these simulations models is that an inverse dynamics approach was the preferred method for predicting movement strategies. Inverse dynamics uses as inputs the joint motions as a function of time and differentiates them twice to yield the accelerations required to calculate joint torques and interaction forces needed to produce the motions. However, the motion functions must be checked for consistency to ensure they stay within geometrical constraints of linked rigid bodies at every time frame. For inverse dynamics with closed topologies, this process is complex and the computation load is quite large. The difficulty in handling closed topologies comes from indeterminacy which a generic property of the inverse dynamics itself [13]. In the present study, the full-body reaching tasks required the subjects to stand firmly on the force plates without any foot movement during the task. Therefore, there exists a closed topology within lower extremities. To avoid the consistency problems of geometrical constraints mentioned above, a one leg model was used in the inverse dynamics calculations used in the optimization and simulation with mass properties doubled for thigh and shank segments. For the experimental data, the same techniques were applied. Given the task constraints, a one leg approximation is acceptable for full-body reaching movements and consistent with our previous work[25].

**Comparison of Simulation and Experiments**

The results of these simulations are in general agreement with the experimental data regarding the displacement of COM and the final postures adopted at different target locations. Perfect fits to the experimental data are not shown in this study because 1) fitting is generally associated with arbitrary parameter adjustments[2], i.e. adjusting some anthropometric parameters may help to improve the predictions [26]; 2) fitting quality may not be sufficient to estimate the validity of a model [27].

Figure 5b illustrates that minimizing Power and COM in the simulation model provided a reasonably close fit to the experimental data. However, it is also clear that there are differences between the simulation and the experimental data regarding location of the end-effector. This is particularly evident for reaches to the middle and low targets. These differences are most likely due to errors associated with motion capture (e.g. skin movement under sensors) and from model constraints (e.g. constraining shoulder translation). Shoulder translation which can dramatically affect location of the end-effector, particularly in forward and overhead reaching tasks [28, 29]. The vertical and anterior-posterior translation could be up to ±3.8cm. However, even with these potential sources of error, both simulation and experimental results capture the major characteristics of the voluntary target reaching movement, i.e. curved path for end-effector, bell-shaped profile of velocity of end-effector and increasing displacement in COM and joint power with lowering the target height.

**Controller**

The minimum jerk solution for end-effector movement based on Euler-Poisson's theorem indicates that the optimal trajectory of end-effector must have the form of a $5^{th}$-order polynomial [3]. Although a $6^{th}$-order polynomial was used in our study, the value of the $6^{th}$-coefficient is very small for minimum error trajectory (e.g. $\|p_6\|_2 = 1.9275 \times 10^{-5}$, $\|p_6\|_2 = 3.9119 \times 10^{-5}$ and $\|p_6\|_2 = 5.4586 \times 10^{-5}$ for high, middle and low target respectively). Thus, the minimum error movement is similar to the movement of minimum jerk of end-effector in terms of symmetric bell-shape velocity profile, sine-shape acceleration profile, and parabolic jerk profile. While for the minimum power and COM trajectory, the value of the $6^{th}$-coefficient is quite big relative to the minimum error trajectory (e.g. $\|p_6\|_2 = 0.0578$, $\|p_6\|_2 = 0.1341$ and $\|p_6\|_2 = 0.4489$ for high, middle and low target respectively). The effects of the $6^{th}$-coefficients of controllers on the minimum power and COM trajectories of end-effector are significant especially for middle and low target reaching movements.

**Optimization Method**

In this study, the Levenberg-Marquardt method with line search was used to find the optimal trajectory of the end-effector in terms of motor task error (end-effector error) or/and certain physiological strategies. It must be emphasized that this method belongs to the category of steepest descent. Therefore, the optimal trajectory is most sensitive to the selected criterion, i.e. the joint angles were calculated in such a way that the increments of joints correspond to the gradient (partial derivatives) of the criterion with respect to the joint angles. The gradient is the minimal change in joint angles that results in unit change of the criterion. This method is consistent with that proposed by Hinton in that each joint is moved autonomously, in proportion to how much moving that joint alone affects the end-effector-target distance. [30]

In addition to using various criteria, any other optimization methods could be used and certainly different optimal trajectories for both end-effector and joint angles can be obtained. Trust region method [31], for instance, can be used to obtain the optimal trajectories for the posture comfort hypothesis if the joint's range of motion is known. A postural comfort hypothesis predicts that joint excursions in multi-joint tasks are in part determined by joint comfort [32-34]. Cruse (1986) pointed out that each joint has an associated discomfort function and that comfort costs influence the movement strategy chosen. The discomfort associated with an individual joint is highest near the joint's biomechanical range limits and lowest for some optimal configuration, which tends to be near the middle of the joint's range of motion [35]. Because certain joints (e.g. knee joint) are near their biomechanical range limits, particularly for reaches to the high target, it is reasonable to assume that the Trust region method would not provide reasonable results for these tasks.

**Criteria**

For full-body voluntary movements, at least two major performance criteria need to be accounted for, i.e. the end-effector must reach the target at the end of movement; and the whole body must maintain balance during the movement (the motion of COM is within the base of support). Other performance criteria, such as minimization of energy, may be necessary, especially for reaches to the low target. Different performance criteria may need to be adopted for different target reaching tasks or even for different periods of the reaching movements [26]. Clearly that more than one performance criterion is required to reasonably predict whole body reaching tasks. In fact, Ferry et al [26] had suggested that even for a simple arm raising task more than one performance criterion may need to be adopted. Additionally, Parnianpour et al [6] reports that even for a movement task consisting of one segment with one degree of freedom may require more than one performance criterion. For end-effector path planning, the minimum motor task error may not be a necessary performance criteria because the location of the end-effector at each moment of time is known [29, 36]. The final boundary condition in Cartesian space also can be put into optimization algorithm constraints. However, the final posture (in joint space) may be required to be known a priori [28]. In principle, path planning shouldn't belong in the voluntary movement category. In path planning, the path in Cartesian space is known; the question is how to solve a redundant inverse kinematics problem. Whereas in voluntary movement the path is unknown, there is no prior knowledge about the movement. In fact, all constraints can be combined into the movement performance criteria. For instance, the joint range can be measured with joint comfort and then combined into the criteria [37].

# Conclusions

In summary, the proposed method in this paper is effective for the simulation of large-scale human skeleton systems, which can reasonable predict whole body reaching movements (i.e. final postures, movement of COM, joint power, and end-effector trajectories etc.). As applied, a combination of several control strategies such as minimizing end-effector error, joint power and COM and using the simple algebraic calculations of inverse dynamics and forward kinematics provided good fits to the experimental data. In the future different cost criteria should be examined and compared with even more complex movement tasks to further elucidate how the CNS plans and executes movements in a kinematics redundant system.

# Competing interests

The authors declare that they have no competing interests.

# Authors' contributions

DS developed the 3D simulation models and contributed substantially to the writing of this manuscript.

JST was responsible for the concept of the reaching task, data collection and analysis of experimental data. JST also contributed to the writing of the manuscript. JST was funded by awards from The National Institutes of Health and Ohio University Post-Doctoral Fellowship Program.

# Acknowledgements

This work was supported by National Institutes of Health Grant grants RO1-HD045512 and from Ohio University Post-Doctoral Fellowship Program Award. We would like to thank Nicole Vander Wiele and Stacey Moenter for their assistance in data collection.

# Figures

**Figure 1 - Model of Optimized Controller**

Block diagram for the optimal model of full body reaching movement. The outputs of controller are the joint angle functions of time. The inverse dynamics and forward kinematics models were used to calculate the physiological measurements and the location of end-effector. Optimization algorithm was used to minimize a certain criterion to produce the optimized parameters for the controller.

**Figure 2 - End-Effector Trajectories**

Trajectories of the end-effector determined by 1) the simulation model using two different cost functions, 2) calculated minimum jerk, and 3) experimental data averaged over 15 subjects are plotted for each target height (gray shadow areas represent the standard error). The left panel is for the high target, middle panel is for middle target, and the right panel is for low target. The trajectories are all remarkably similar for target distance, velocity, and acceleration. The largest differences emerge for jerk (bottom row) where the experimental data are not well fit by any of the models. The top row indicates the distance between the end-effector and target location.

**Figure 3 - End-Effector Pats**

The path of the end-effector determined by the different simulation methods and the experimental data (as described above) for each plane and target height are plotted. The paths for plotted for **A**. sagittal plane all target heights **B**. frontal plane all target heights **C**. transverse plane for high target **D**. transverse plane for middle target E. transverse plane for low target.

**Figure 4 - Center-of-Mass Movement**

Simulation and experimental motion of whole body COM in horizontal plane for high (top panel), middle (middle panel) and low target (bottom panel) respectively. Gray shadow areas represent bi-directional standard error. Large variations of AP displacements are shown within subjects.

**Figure 5 - Posture Adopted at Target Contact**

Comparison of final postures between simulations with different cost functions (upper panel) and between simulation (min power & COM) and observed posture (lower panel). Left panel is for high target, middle panel is for middle target, and right panel is for low target.

**Figure 6 – Total Joint Power**

Comparison of the total joint power (sum of absolute each joint power) from simulation with two kinds of criteria for high (top panel), middle (middle panel) and low target (bottom panel) respectively.

# Tables

## Table 1 - Input Joint Data

Range of motion and viscoelastic coefficients of joints.

| Joint | Plane | Limitation (deg)[e] | | Viscoelastic coefficients | |
| --- | --- | --- | --- | --- | --- |
| | | upper | lower | Stiffness K (N.m.deg$^{-1}$) | Damper B (N.m.deg$^{-1}$.s) |
| Ankle | Flexion/extension | 54.3 | -12.2 | 1/6[a] | |
| | Int/external Rotation | 0.01 | -0.01 | | |
| | Add/abduction | 19.2 | -19.2 | 1/15[a] | |
| Knee | Flexion/extension | 141.2 | -0.01 | 1/20[a] | |
| | Int/external Rotation | 0.01 | -0.01 | | |
| | Add/abduction | 0.01 | -0.01 | | |
| Hip | Flexion/extension | 12.1 | -121.3 | 1/3[a] | |
| | Int/external Rotation | 44.2 | -44.2 | | |
| | Add/abduction | 25.6 | -25.6 | 1[a] | |
| L Shoulder | Flexion/extension | 62 | -167 | 0.192[d] | 0.014[d] |
| | Int/external Rotation | 69 | -104 | 0.192[d] | 0.014[d] |
| | Add/abduction | 184 | -0.01 | 0.192[d] | 0.014[d] |
| L Elbow | Flexion/extension | 0.3 | -140.5 | 0.1571[d] | 0.0122[d] |
| | Int/external Rotation | 81.1 | -75 | 0.1571[d] | 0.0122[d] |
| | Add/abduction | 0.01 | -0.01 | | |
| L Wrist | Flexion/extension | 35.3 | -21.1 | 0.1047[d] | 0.0105[d] |
| | Int/external Rotation | 0.01 | -0.01 | 0.1047[d] | 0.0105[d] |
| | Add/abduction | 74 | -74.8 | 0.1047[d] | 0.0105[d] |
| Cervical | Flexion/extension | 141 | -141 | 0.25[b] | |
| | Int/external Rotation | 93 | -93 | 0.42[b] | |
| | Add/abduction | 172 | -172 | 0.33[b] | |
| Thorax | Flexion/extension | 27 | -27 | 0.25[c] | |
| | Int/external Rotation | 21 | -21 | 0.42[c] | |
| | Add/abduction | 4 | -4 | 0.33[c] | |
| Lumbar | Flexion/extension | 43 | -43 | 0.25[c] | |
| | Int/external Rotation | 19 | -19 | 0.42[c] | |
| | Add/abduction | 8 | -8 | 0.33[c] | |
| R Shoulder | Flexion/extension | 62 | -167 | 0.192[d] | 0.014[d] |
| | Int/external Rotation | 69 | -104 | 0.192[d] | 0.014[d] |
| | Add/abduction | 0.01 | -184 | 0.192[d] | 0.014[d] |
| R Elbow | Flexion/extension | 0.3 | -140.5 | 0.1571[d] | 0.0122[d] |
| | Int/external Rotation | 75 | -81.1 | 0.1571[d] | 0.0122[d] |
| | Add/abduction | 0.01 | -0.01 | | |
| R Wrist | Flexion/extension | 35.3 | -21.1 | 0.1047[d] | 0.0105[d] |
| | Int/external Rotation | 0.01 | -0.01 | 0.1047[d] | 0.0105[d] |
| | Add/abduction | 74.8 | -74 | 0.1047[d] | 0.0105[d] |

[a] calculated from Amankwah *et al*. 2004.
[b] adopted from De Jager, 1996.
[c] adopted from Moroney *et al* 1988.
[d] adopted from Gomi and Osu 1998.

**Table 2 - Computational Costs**

Costs of optimal control strategies and CPU times for each target location are shown below.

| Target location | Measurement | Optimal control strategy | | | |
|---|---|---|---|---|---|
| | | min Error | min Power | min COM | min Power & COM |
| High | Total power squared $\int_0^{t_f} (\tau \times \dot{\theta})^T (\tau \times \dot{\theta}) dt$ (J$^2$) | 3436 | 320.2 | 285.7 | 130 |
| | Final COM squared $x_{Cf}^T x_{Cf}$ (m$^2$) | 0.03509 | 0.01309 | 0.000453 | 0.00784 |
| | CPU time (hour) | 1.84h[a] | 3.90h[b] | 1.96h[b] | 5.43h[a] |
| Middle | Total power squared | 5608 | 1023 | 1672 | 353.4 |
| | Final COM squared | 0.05101 | 0.02597 | 0.00047 | 0.003058 |
| | CPU time | 1.13h[b] | 2.37 h[b] | 2.66 h[b] | 2.73h[b] |
| Low | Total power squared | 11167 | 8638 | 13940 | 3705 |
| | Final COM squared | 0.09823 | 0.0917 | 0.00155 | 0.01258 |
| | CPU time | 2.33h[b] | 4.76 h[a] | 2.99 h[b] | 3.11h[a] |

[a] Desktop computer: Intel Xeon, 3.20GHz and 3.19GHz, 2GB of RAM, Windows XP, Matlab 2006b

[b] Laptop computer: Intel Pentium M, 1.86GHz, 1GB of RAM, Windows XP, Matlab 2006b

**Table 3 - COM comparisons**

Final COM displacements (mm) derived from simulation models are compared to experimental results for each target location. Mean values from experimental data (± SEM) are also presented.

|  |  | High target | Middle target | Low target |
|---|---|---|---|---|
| Min task error | Anterior-posterior | 172.3* | 219.7* | 311.2* |
|  | Mediolateral | 73.6 * | 52.5* | 37.3 |
| Min Power | Anterior-posterior | 113.6* | 160.3* | 299* |
|  | Mediolateral | 13.3* | 16.8* | 47.9* |
| Min Com | Anterior-posterior | 2* | 17.9* | 39.2* |
|  | Mediolateral | 21.2* | 12.3* | 3.7(ns) |
| Min Power & COM | Anterior-posterior | 49.1 (ns) | 60.2 (ns) | 112 (ns) |
|  | Mediolateral | 2.1 (ns) | 5 (ns) | 5 (ns) |
| Experiment | Anterior-posterior | 58.7±17.9 | 67.8±20.8 | 86.6±21.2 |
|  | Mediolateral | -8.5±6.7 | -7.1±6.8 | 3.5±7.1 |

* indicates p<.05

**Table 3 - Total energy comparisons**

Comparison of total energy (J) between model simulations and their corresponding experimental results of all joints are shown for each target location. Mean values for experimental data (± SEM) are also presented.

|  | High target | Middle target | Low target |
|---|---|---|---|
| min Error | 52.98* | 61.64 (ns) | 115.9 * |
| min Power | 21.06 * | 35.46 * | 104.5 * |
| min COM | 29.03 * | 56.24 * | 155.4 * |
| min Power & COM | 15.21 * | 32.59 * | 90.38 * |
| Experiment | 27.6±12.44 | 50.86±21.47 | 98.75±61.33 |

Figure 1

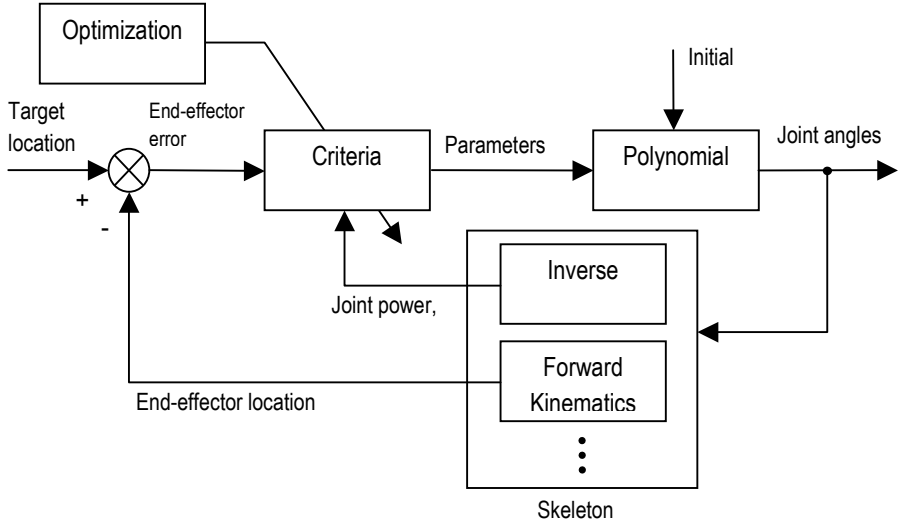

Figure 2

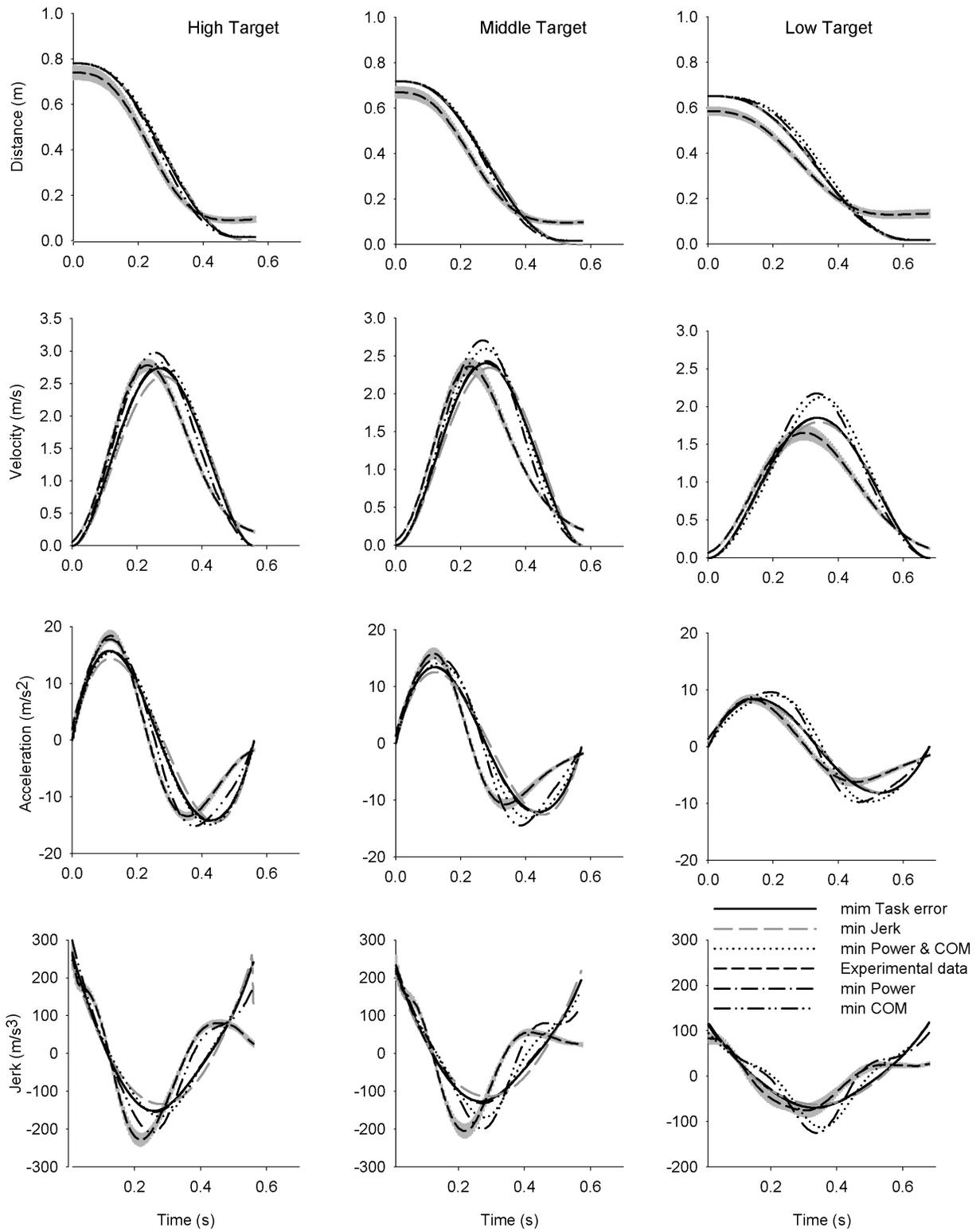

Figure 2

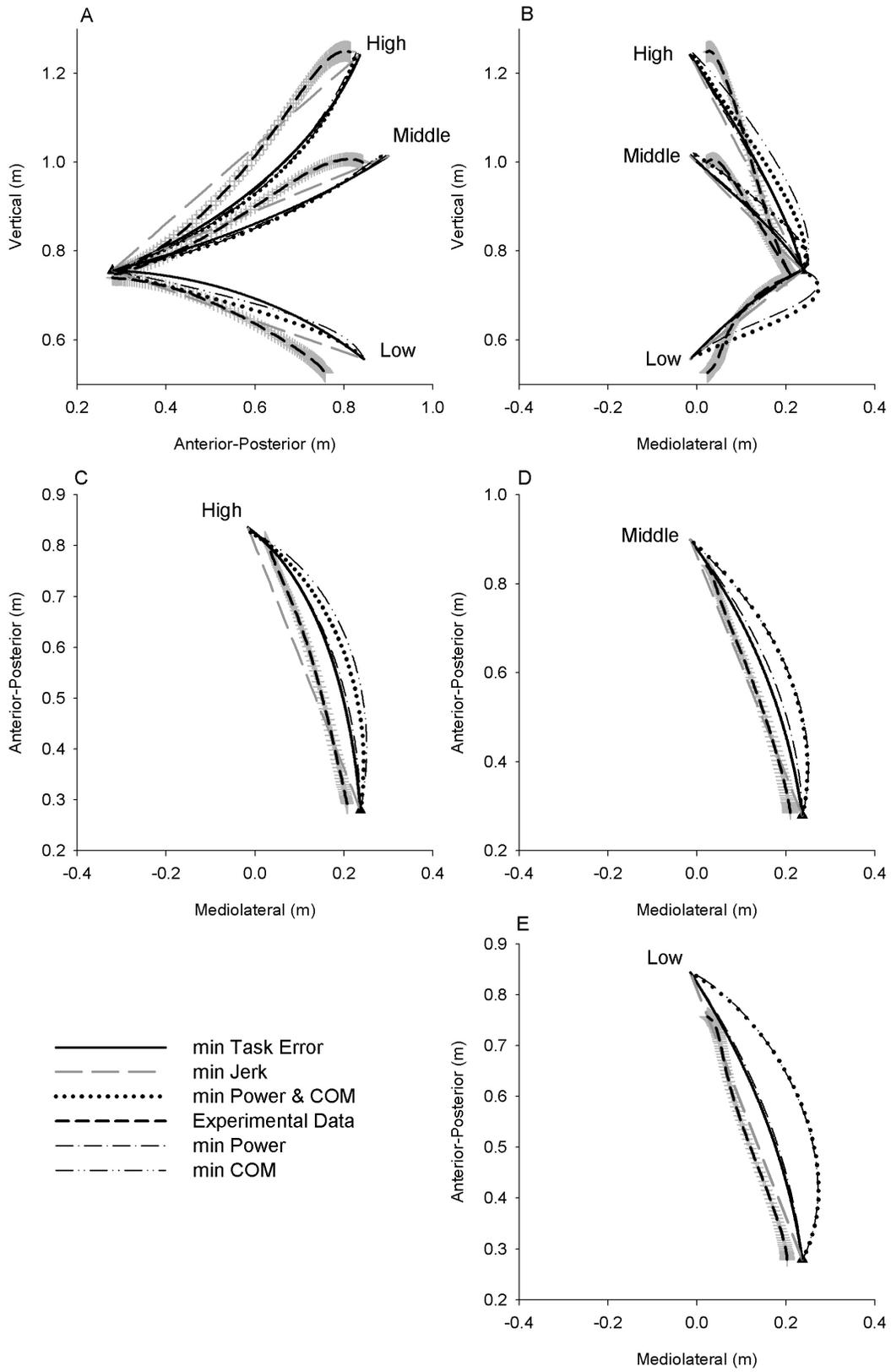

Figure 3

Figure 4

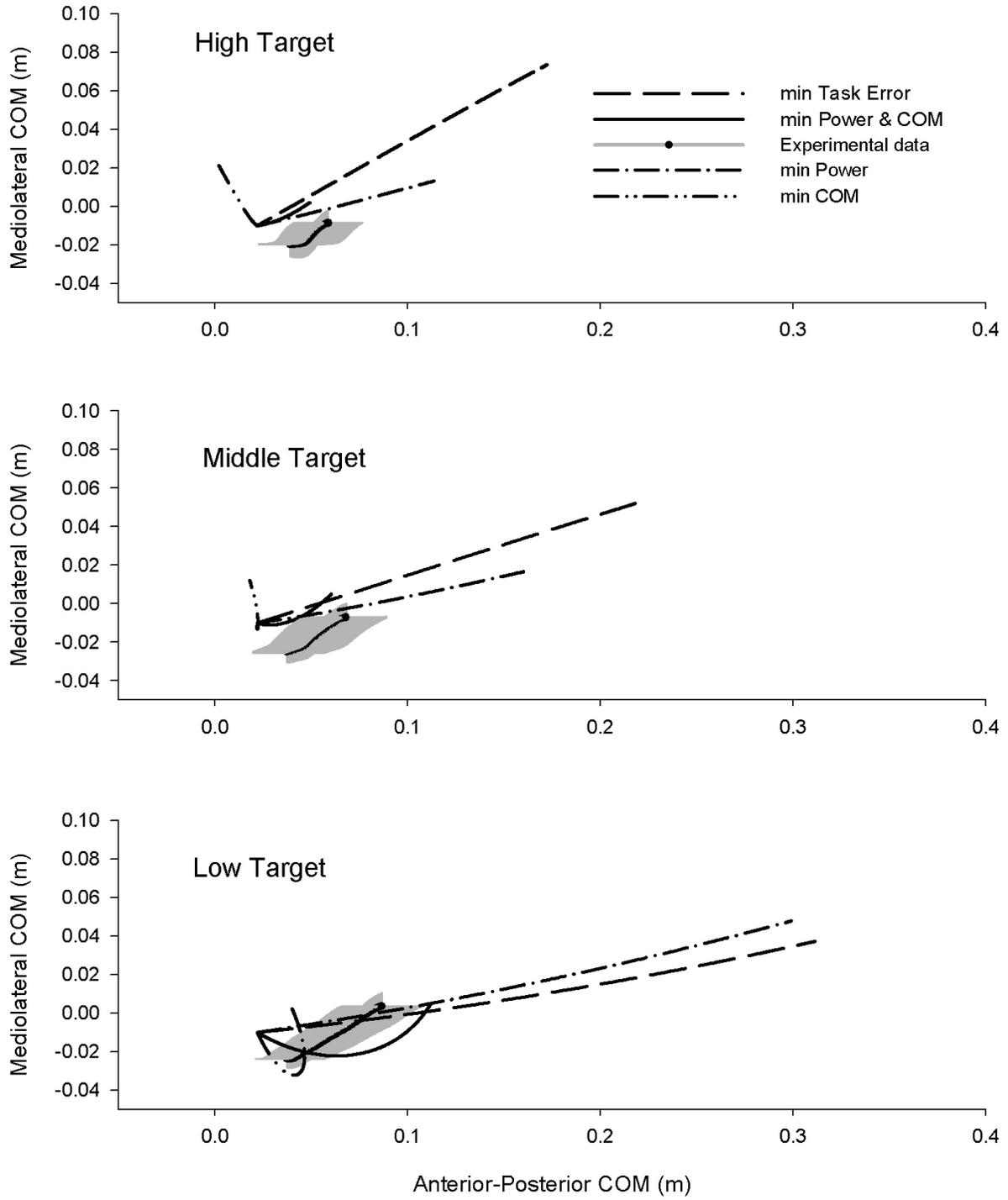

Figure 5

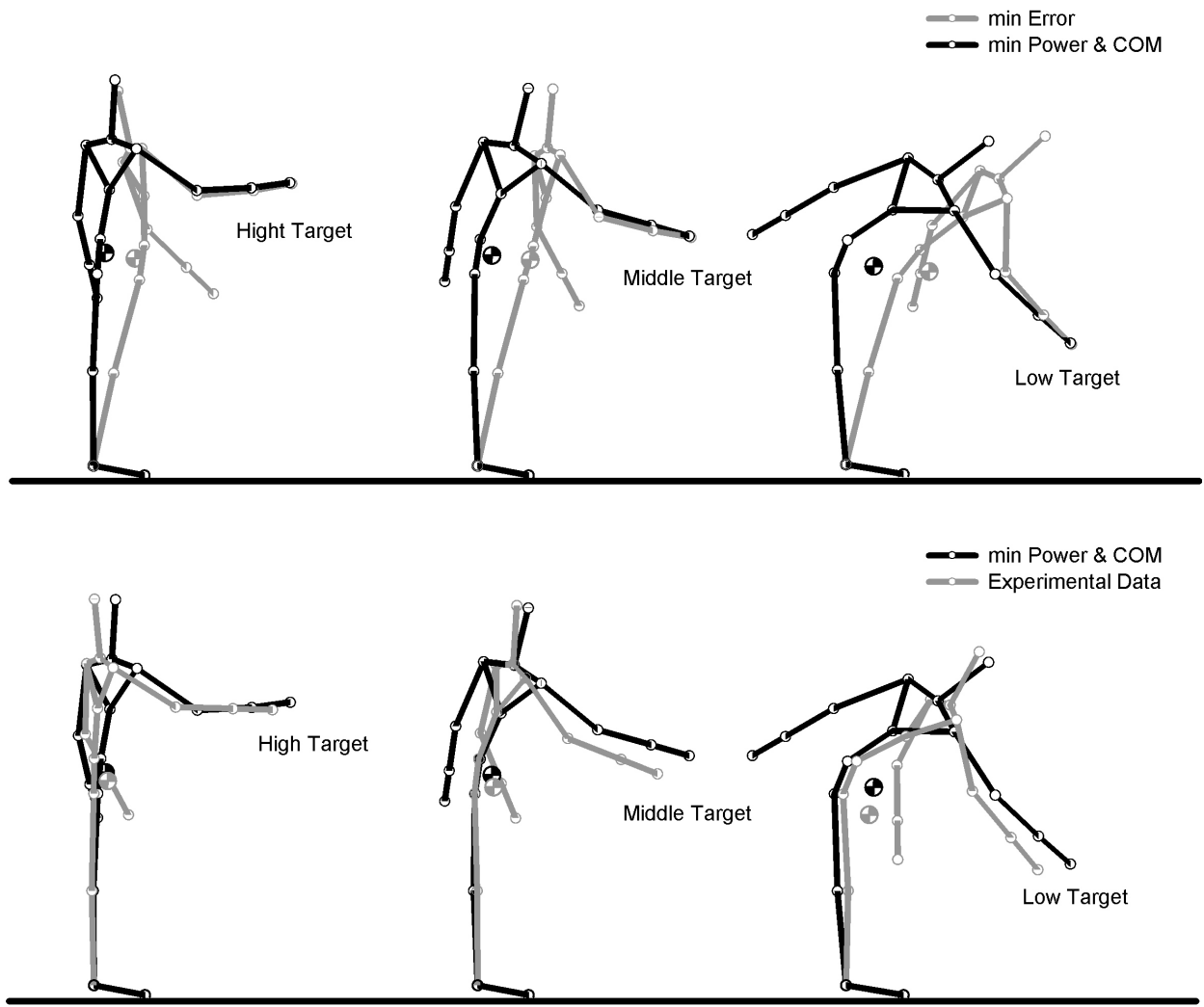

Figure 6

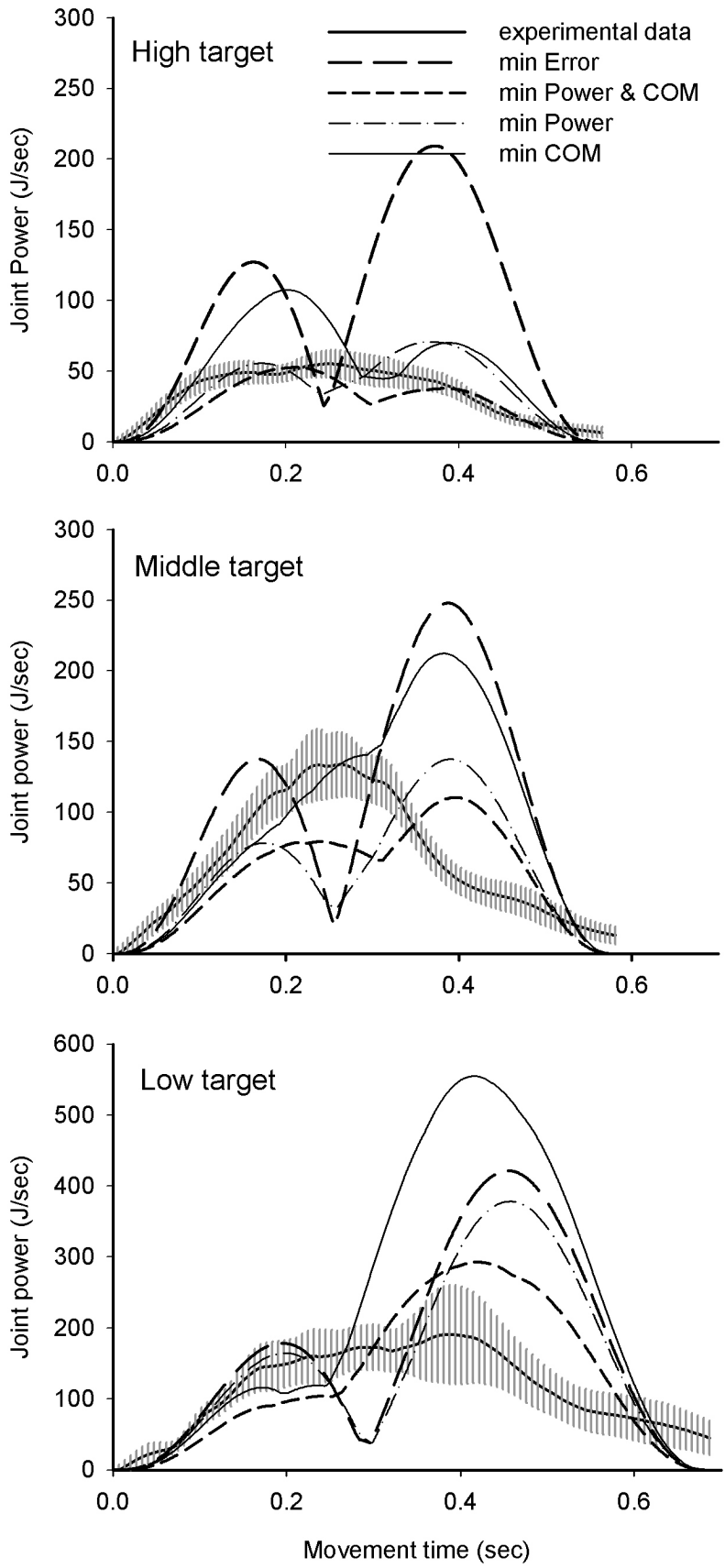